\documentclass{article}

\usepackage{arxiv}
\usepackage[utf8]{inputenc} 
\usepackage[T1]{fontenc}    
\usepackage{cite}
\usepackage{amsmath,amssymb,amsfonts}
\usepackage{algorithmic}
\usepackage{graphicx}
\usepackage{textcomp}
\usepackage{marvosym}
\usepackage{url}
\usepackage{xcolor}
\usepackage{{booktabs}}
\usepackage{tcolorbox}
\usepackage{subcaption}

\title{Evaluating LLMs for Text-to-SQL Generation With Complex SQL Workload}

\usepackage{authblk}

\author[1]{%
	Limin Ma\thanks{\texttt{limin.ma@ontariotechu.net}}}%
\author[1]{%
	Ken Pu\thanks{\texttt{ken.pu@ontariotechu.ca}}}%
\author[2]{%
	Ying Zhu\thanks{\texttt{ken.pu@ontariotechu.ca}}}%
\affil[1]{Faculty of Science, Ontario Tech University}
\affil[2]{Faculty of Business and IT, Ontario Tech University}

\renewcommand{\shorttitle}{\textit{arXiv} Submission}

\begin{document}

\maketitle


\begin{abstract}
    This study presents a comparative analysis of the a complex SQL benchmark, TPC-DS, with two existing text-to-SQL benchmarks, BIRD and Spider. Our findings reveal that TPC-DS queries exhibit a significantly higher level of structural complexity compared to the other two benchmarks. This underscores the need for more intricate benchmarks to simulate realistic scenarios effectively. To facilitate this comparison, we devised several measures of structural complexity and applied them across all three benchmarks. The results of this study can guide future research in the development of more sophisticated text-to-SQL benchmarks.

    We utilized 11 distinct Language Models (LLMs) to generate SQL queries based on the query descriptions provided by the TPC-DS benchmark. The prompt engineering process incorporated both the query description as outlined in the TPC-DS specification and the database schema of TPC-DS. Our findings indicate that the current state-of-the-art generative AI models fall short in generating accurate decision-making queries. We conducted a comparison of the generated queries with the TPC-DS gold standard queries using a series of fuzzy structure matching techniques based on query features. The results demonstrated that the accuracy of the generated queries is insufficient for practical real-world application.
\end{abstract}


\section{Introduction}
The task of generating SQL queries from natural language (NL) has been a long-standing problem in the field of natural language processing (NLP) and databases. The task is important as it can enable non-expert users to interact with databases without having to learn SQL. The task is also important for database administrators who can use the generated SQL queries as a starting point for further optimization. The task is also important for data scientists who can use the generated SQL queries to analyze data and generate insights.

In recent years, large language models (LLMs) have shown impressive performance on a wide range of NLP tasks. LLMs are pre-trained on large amounts of text data and fine-tuned on specific tasks. LLMs have been shown to achieve state-of-the-art performance on a wide range of NLP tasks, including text-to-SQL generation. However, the task of generating SQL queries from NL is challenging due to the complex structure of SQL queries and the need for accurate decision making.

In the effort to evaluate the performance of LLMs on the task of generating SQL queries from NL, two major benchmarks have been proposed: BIRD and Spider. BIRD is a benchmark for text-to-SQL generation that consists of over 12,000 NL questions and their corresponding SQL queries. Spider is a benchmark for text-to-SQL generation that consists of 10,000 NL questions corresponding 5693 SQL queries. Both benchmarks have been widely used to evaluate the performance of LLMs on the task of generating SQL queries from NL.

In this study, we present a comparative analysis of the TPC-DS benchmark \cite{poess2000new} with the BIRD \cite{li2024can} and Spider \cite{yu2018spider} benchmarks. The TPC-DS benchmark is a widely used benchmark for evaluating the performance of database systems. The benchmark consists of a set of complex SQL queries that simulate real-world decision-making queries. Our findings reveal that TPC-DS queries exhibit a significantly higher level of structural complexity compared to the BIRD and Spider benchmarks. This underscores the need for more intricate benchmarks to simulate realistic scenarios effectively.  Furthermore, we utilized 11 distinct LLMs to generate SQL queries based on the query descriptions provided by the TPC-DS benchmark. The prompt engineering process incorporated both the query description as outlined in the TPC-DS specification and the database schema of TPC-DS. Our findings indicate that the current state-of-the-art generative AI models fall short in generating accurate decision-making queries. We conducted a comparison of the generated queries with the TPC-DS gold standard queries using a series of fuzzy structure matching techniques based on query features. The results demonstrated that the accuracy of the generated queries is insufficient for practical real-world application.

\begin{figure}[t]
    \centering
    \includegraphics[width=0.8\textwidth]{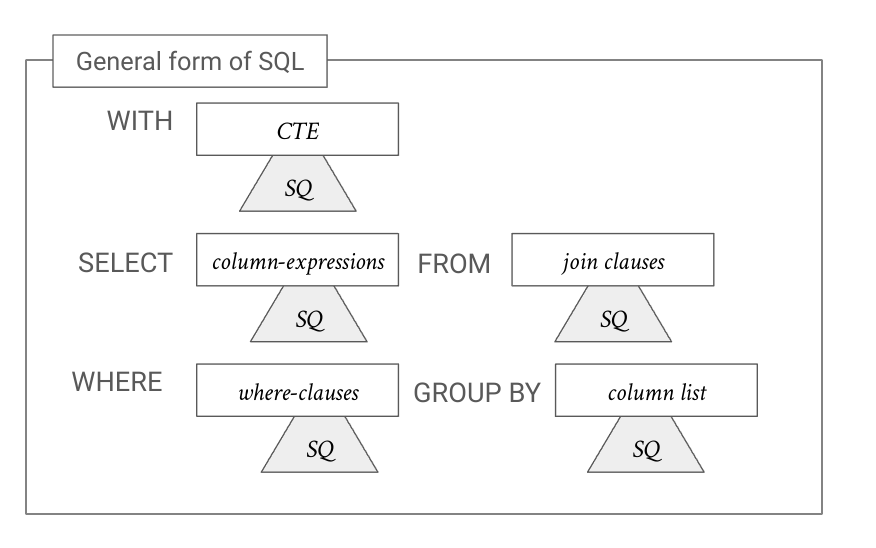}
    \caption{The general form of a SQL query.  Note all triangles
    labeled as {\bf\it SQ} are possible occurrences of sub-queries.}
    \label{fig:sql-form}
\end{figure}

\section{Related Work}

The field of text-to-SQL has seen significant progress owing to the integration of large language models (LLMs) \cite{naveed2023comprehensive,bae2024ehrxqa}.  Innovative techniques have been proposed for improving the accuracy and efficiency of SQL query generation from natural language inputs. We discuss several key works in this domain, and present their contributions and how they relate to our work.

Owda et al. \cite{owda2011information} present an early system that incorporates Information Extraction (IE) techniques into an Enhanced Conversation-Based Interface of Relational Databases (C-BIRD) to generate dynamic SQL queries. Their approach allows conversational agents to interact with users in natural language, generating SQL queries without any user requirement of prior SQL knowledge.  Since then, LLM-based solutions have dominated the field of text-to-SQL.  Furthermore, two major benchmarks, BIRD and Spider, have been proposed to evaluate the performance of LLMs on the task of generating SQL queries from NL.  
Yu et al. \cite{yu2018spider} introduce Spider, a large-scale, complex, and cross-domain semantic parsing and text-to-SQL dataset. Spider includes 10,181 questions and 5,693 unique complex SQL queries across 200 databases covering 138 different domains, requiring models to generalize well to both new SQL queries and database schemas. This work is foundational in the text-to-SQL field, demonstrating the challenges in handling complex and diverse SQL queries. Li et al. \cite{li2024can} introduce BIRD, a large-scale benchmark for text-to-SQL tasks aimed at bridging the gap between academic research and real-world applications. BIRD benchmark measures both the accuracy and efficiency of text-to-SQL models, providing a comprehensive evaluation of model performance.

A number of text-to-SQL models and techniques have been proposed and were heavily based on the BIRD and Spider benchmarks.  Li et al. \cite{li2019comprehensive} introduce the Fuzzy Semantic to Structured Query Language (F-SemtoSql) neural approach, designed to tackle the complex and cross-domain task of text-to-SQL generation.  Pourreza and Rafiei \cite{pourreza2024din} propose DIN-SQL, a method that improves text-to-SQL performance by decomposing the task into smaller sub-tasks and using in-context learning with self-correction.
Chen et al. \cite{chen2024open} present the Open-SQL Framework, designed to enhance text-to-SQL performance on open-source LLMs. They introduce novel strategies for supervised fine-tuning and the {\em openprompt} strategy for effective question representation. 
Li et al. \cite{li2023resdsql} propose ResdSQL, a framework that decouples schema linking and skeleton parsing for text-to-SQL tasks. By addressing the complexity of parsing schema items and SQL skeletons separately, their method improves the accuracy of SQL query generation.

\section{Features And Complexity Measures of SQL Queries}
\label{sec:features}

Consider a SQL query $Q$ as shown in Figure~\ref{fig:sql-form}.
Given the nested structures of SQL, $Q$ can be a complex self-referencing tree of SQL queries.  In this section, we will define a number of features that can be used to characterize the structural complexity of SQL queries.  We argue that the bag-valued features are useful in performing semantic comparison of pairs of SQL queries.

Let $\mathbf{SQ}(Q)$ be all sub-queries in $Q$
regardless of where it appears in $Q$.  We define a collection of features derived from $Q$ and its sub-queries.  These features are divided into two categories: bag-valued features and numeric features.  Namely, a feature $F$
is a mapping:
$$ F_\mathrm{bag}: \mathbf{SQL}\rightarrow\mathbf{Bags}
\quad\mathrm
F_\mathrm{count}: \mathbf{SQL}\to\mathbb{N} $$

We note that any bag feature $F$ is also naturally a numeric feature $F^{\#}$:
$F^{\#}(Q) = |F(Q)|$. 

\subsection{Bag-valued Features}
\noindent {\bf Columns}: the set of distinct columns included in the select {\em column expressions} of $Q$ and its subqueries.
\begin{align*}
    \mathrm{Cols}(Q) =& \{c\in\mathrm{Columns}: c\in\mathrm{SelectExpressions}(Q)\} \\
    &\cup\bigcup_{Q'\in\mathrm{SQ}(Q)}\mathrm{Cols}(Q')
\end{align*}

\noindent{\bf Relations}: $\mathrm{Tables}(Q)$ the set of distinct relations in the SQL query.  This is computed as:
\begin{align*}
    \mathrm{Relations}(Q) = & \{r\in\mathrm{Relations}: r\in\mathrm{FromClause}(Q)\} \\
    & \cup\bigcup_{Q'\in\mathrm{SQ}(Q)}\mathrm{Relations}(Q')
\end{align*}

\noindent{\bf Where predicates}: $\mathrm{WHERE}(Q)$ the set of distinct where basic predicates in the SQL query and its subqueries.  Each basic predicate is given in the form of
$$ \left<\mathrm{pred}, \mathrm{expr}_i, \dots\right> $$
where $\mathrm{pred}$ is the boolean operator supported by SQL, and $\mathrm{expr}_i$ is an expression over column names, constants, and functions.  All columns aliases are normalized to the canonical form.

\noindent{\bf JOINs}: $\mathrm{JOIN}(Q)$ is the set of joins of $Q$ is defined to be pairs of physical relations $(t, t')$ that participate in a JOIN in the {\em logical} query plan of $Q$ and its subqueries.

\noindent{\bf Aggregation}: $\mathrm{Agg}(Q)$ is the set of aggregated columns in $Q$ and its subqueries.  Each aggregated column is given in the form of $\left<\mathrm{agg}, \mathrm{expr}, \mathrm{groupby\ columns}\right>$ where $\mathrm{agg}$ is the aggregation function (e.g., SUM, AVG, COUNT, etc.) and $\mathrm{expr}$ is an expression over column names, constants, and functions.

\noindent {\bf Functions}: $\mathrm{Func}(Q)$ is the set of SQL functions that appear in $Q$ and its subqueries.

\subsection{Numeric Features}

\noindent{\bf Common Table Expressions}: $\mathrm{CTE}(Q)$ is the number of common-table-expressions defined by $Q$.  We use CTE count as an indicator of the structural complexity and readability of the query $Q$.

\noindent{\bf Sub-queries}: we measure $\|\mathrm{SQ}(Q)\|$, the number of subqueries of $Q$ as an indicator of the structural complexity of $Q$.

\section{TPC-DS As A Complex Text-to-SQL Benchmark}

\begin{figure}[t]
    \centering
    \includegraphics[width=0.8\textwidth]{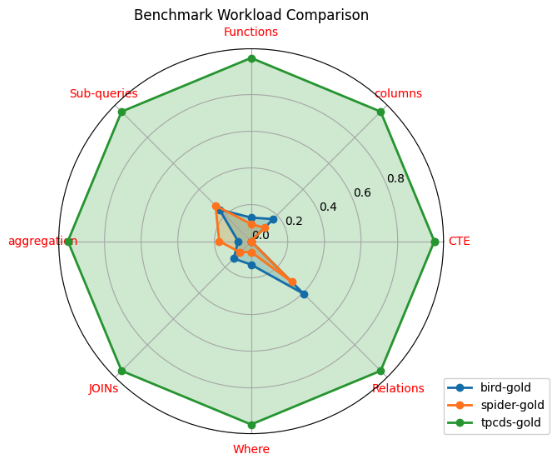}
    \caption{Feature based comparison of SQL benchmarks}
    \label{fig:radar}
\end{figure}

\begin{figure*}[t]
    \centering
    \begin{subfigure}[b]{0.45\textwidth}
        \includegraphics[width=\textwidth]{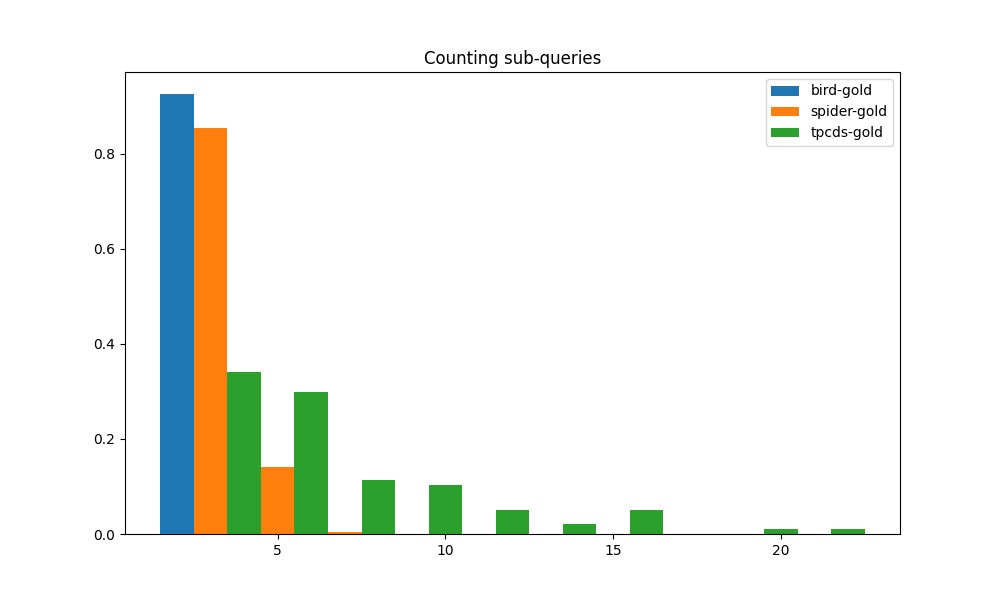}
        \caption{$\mathrm{SQ}(Q)$}
        \label{fig:subquery}
    \end{subfigure}
    \begin{subfigure}[b]{0.45\textwidth}
        \includegraphics[width=\textwidth]{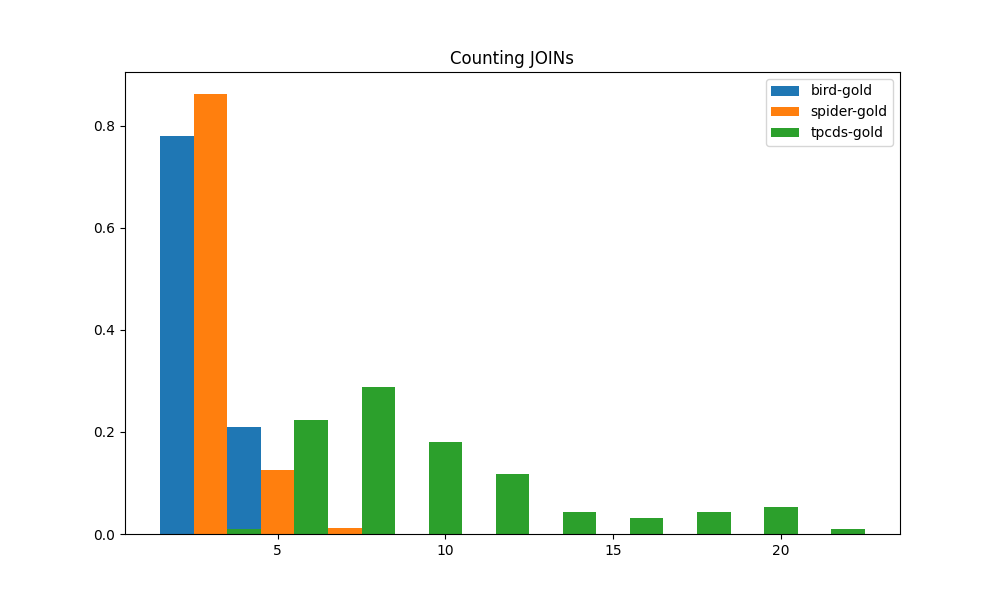}
        \caption{$\mathrm{JOIN}(Q)$}
        \label{fig:join}
    \end{subfigure}

    \begin{subfigure}[b]{0.45\textwidth}
        \includegraphics[width=\textwidth]{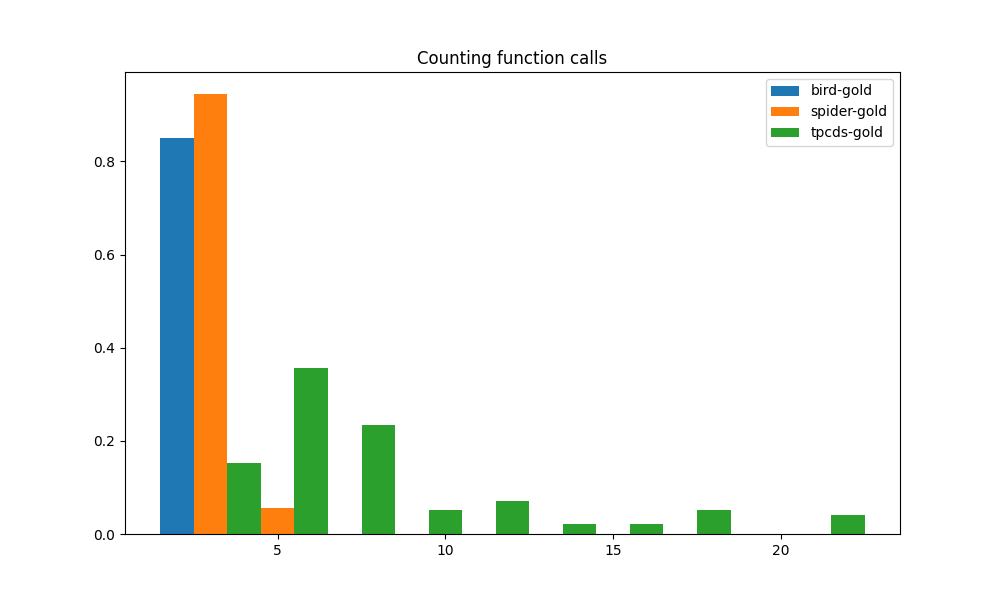}
        \caption{$\mathrm{Func}(Q)$}
        \label{fig:func-call}
    \end{subfigure}
    \begin{subfigure}[b]{0.45\textwidth}
        \includegraphics[width=\textwidth]{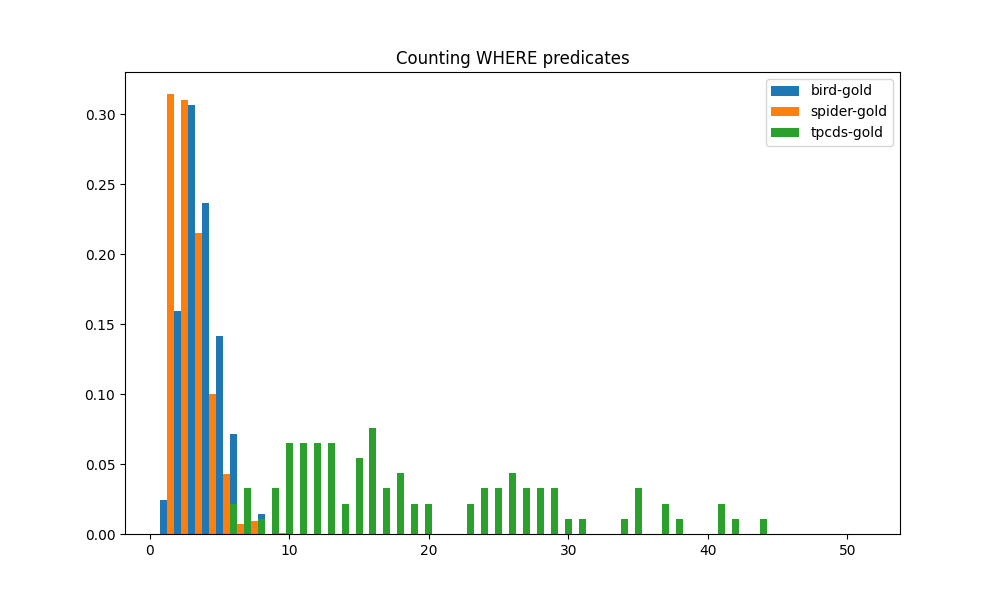}
        \caption{$\mathrm{WHERE}(Q)$}
        \label{fig:where}
    \end{subfigure}

    \begin{subfigure}[b]{0.45\textwidth}
        \includegraphics[width=\textwidth]{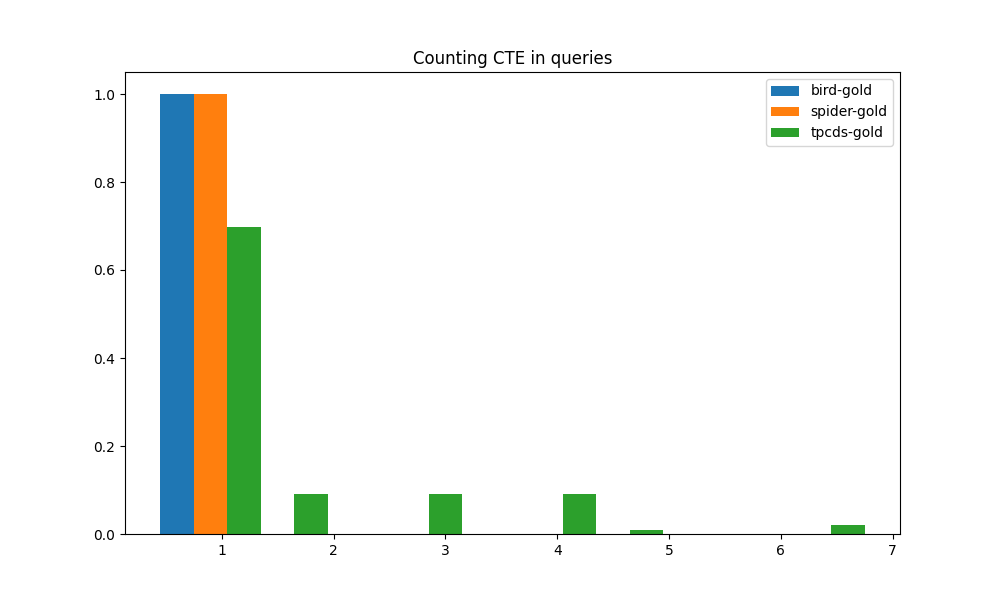}
        \caption{$\mathrm{CTE}(Q)$}
        \label{fig:cte}
    \end{subfigure}
    \begin{subfigure}[b]{0.45\textwidth}
        \includegraphics[width=\textwidth]{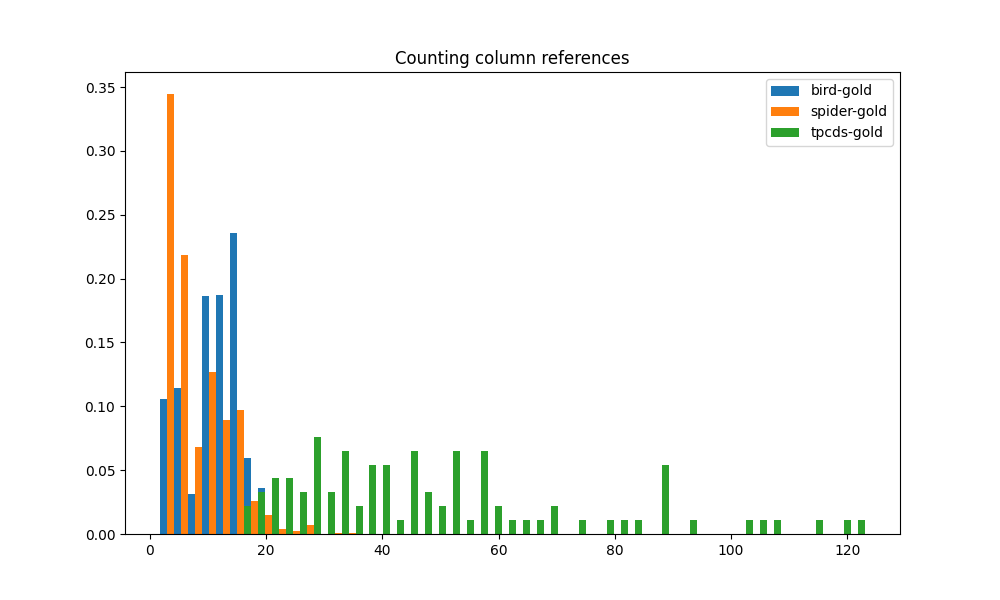}
        \caption{$\mathrm{Cols}(Q)$}
        \label{fig:column}
    \end{subfigure}
    
    \caption{Feature based comparison of benchmarks}
\end{figure*}

We compare the complexity of WHERE clauses in the SQL queries from the three SQL benchmarks.  Complexity of the WHERE clause is measured by the number of basic predicates in the WHERE clause.  A basic predicate is one that has a single condition, such as \texttt{year = 1998}.  Compound predicates comprising multiple conditions, such as \texttt{year = 1998 AND ss\_quantity BETWEEN 81 AND 100}, are counted as multiple basic predicates.  The counts of basic predicates in WHERE clauses for the three benchmarks are plotted in a histogram in Fig.~\ref{fig:where}.  While the BIRD and SPIDER benchmarks are similar in their distribution of WHERE clause predicate counts, the TPC-DS benchmark clearly exhibits substantially greater WHERE clause complexity in its SQL queries.  TPC-DS has such greater WHERE clause complexity than the other two benchmarks that its distribution of the counts hardly overlaps with the others.  Almost all TPC-DS queries have multiple times the WHERE predicate counts than all the queries from the other two benchmarks.

\begin{table*}[btbp]
    \centering
    {\small
\begin{tabular}{lrrrrrrrr}
\toprule
 & CTE & Selected Columns & Functions & Nested Select & Aggregation & joinCount & Where & Joins \\
dataset &  &  &  &  &  &  &  &  \\
\midrule
tpcds-gold & 1.00 & 1.00 & 1.00 & 1.00 & 1.00 & 1.00 & 1.00 & 1.00 \\
gemini-1.5 & 2.38 & 0.79 & 0.62 & 1.01 & 0.95 & 0.91 & 0.55 & 0.87 \\
gpt-4 & 1.11 & 0.77 & 0.63 & 0.69 & 0.80 & 0.73 & 0.53 & 0.90 \\
codestral & 0.19 & 0.67 & 0.69 & 0.53 & 0.63 & 0.63 & 0.53 & 0.90 \\
mixtral-8x22b & 1.99 & 0.96 & 0.70 & 1.04 & 1.06 & 0.96 & 0.67 & 0.90 \\
llama3-70b & 2.03 & 0.81 & 0.75 & 1.37 & 1.02 & 1.04 & 0.67 & 0.94 \\
llama3-8b & 0.51 & 0.63 & 0.63 & 0.67 & 0.65 & 0.59 & 0.52 & 0.84 \\
codellama-7b & 0.03 & 0.39 & 0.29 & 0.28 & 0.32 & 0.24 & 0.25 & 0.55 \\
\bottomrule
\end{tabular}
}

    \caption{Feature based comparison of generated SQL queries}
    \label{table:gen-table}
\end{table*}

\begin{table*}[t]
    \centering
    \begin{tabular}{lrrrrrrr|r}
\toprule
feature & Tables & Columns & Where Predicates & Constants & Functions & Aggregation & Joins & Average \\
model &  &  &  &  &  &  &  &  \\
\midrule
gemini-1.5 & 0.74 & {\color{blue}\bf 0.26} & 0.04 & 0.27 & {\color{blue}\bf 0.67} & {\color{blue}\bf 0.29} & 0.06 & {\bf \color{blue}0.33} \\
gpt-4 & {\color{blue}\bf 0.76} & 0.22 & 0.02 & 0.32 & 0.65 & 0.25 & 0.04 & 0.32 \\
codestral & 0.67 & 0.22 & 0.04 & 0.27 & 0.62 & 0.23 & 0.05 & 0.30 \\
mixtral-8x22b & 0.69 & 0.17 & 0.02 & {\color{blue}\bf 0.30} & 0.60 & 0.24 & 0.06 & 0.30 \\
mistral-large & 0.66 & 0.19 & {\color{blue}\bf 0.06} & 0.22 & 0.54 & 0.24 & 0.08 & 0.28 \\
llama3-70b & 0.63 & 0.13 & 0.04 & 0.25 & 0.60 & 0.12 & {\color{blue}\bf 0.12} & 0.27 \\
codellama-70b & 0.36 & 0.12 & 0.03 & 0.13 & 0.44 & 0.11 & 0.11 & 0.18 \\
llama3-8b & 0.36 & 0.05 & 0.01 & 0.16 & 0.47 & 0.05 & 0.07 & 0.17 \\
codellama-13b & 0.28 & 0.08 & 0.02 & 0.10 & 0.34 & 0.05 & 0.07 & 0.13 \\
codellama-34b & 0.24 & 0.06 & 0.01 & 0.09 & 0.32 & 0.05 & 0.07 & 0.12 \\
codellama-7b & 0.20 & 0.05 & 0.01 & 0.05 & 0.32 & 0.04 & 0.04 & 0.10 \\
\bottomrule
\end{tabular}

    \caption{Similarity comparison of generated queries with TPC-DS gold queries}
    \label{table:similarity}
\end{table*}

\begin{figure}[htbp]
    \centering
    \includegraphics[width=0.8\textwidth]{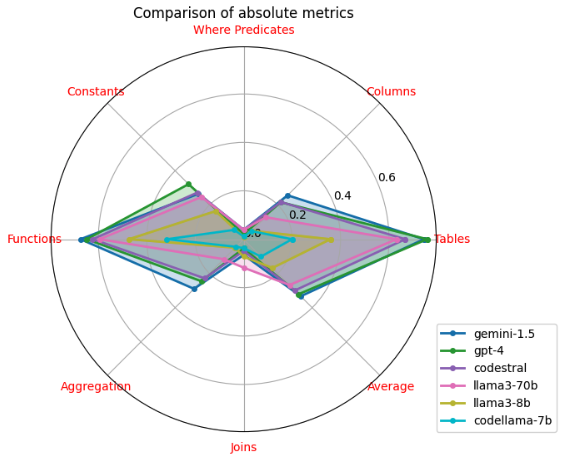}
    \caption{Feature based similarity of generated queries with TPC-DS gold queries}
    \label{fig:gen-radar}
\end{figure}

Another metric used to measure the complexity of SQL queries is the number of Common Table Expressions (CTE).  The role of a CTE is to minimize repetition of subqueries as well as making the overall SQL statement more structured and thus easier to maintain.  We count the number of CTE in each query from the three benchmarks, and plot the histogram in Fig.~\ref{fig:cte}.  It can be seen that both BIRD and SPIDER workloads do not require CTE due to their low degree of semantic complexity.  In contrast, TPC-DS includes heavy usage of CTE in the gold queries.

We also count the number of distinct columns referenced in SQL queries.  The number of distinct columns involved in a query is an indicator of the semantic complexity of that query.  A column is included in the count whether it appears in \texttt{SELECT} projections, in \texttt{JOIN} conditions, or in \texttt{WHERE} predicates.  The distributions of the column counts for the benchmarks are shown in Fig.~\ref{fig:column}.  This graph is similar to the one for the count of WHERE clause predicates.  Again, the number of columns used in the gold queries of TPC-DS is, in most cases, an order of magnitude larger than the number of columns in BIRD and SPIDER queries.  The overlap is minimal, which means that the vast majority of TPC-DS workload reference a much larger set of columns than BIRD and SPIDER workloads.  This is another indication of the significantly higher query complexity of TPC-DS.

In addition to the above, we also consider three more metrics towards our evaluation of overall query complexity.
SQL queries often include calls to functions that perform data transformation.  These functions include scalar and aggregation functions.  They contribute to the overall query complexity.  We therefore count the number of expressions that contain function calls, whether they are part of a SELECT projection, a WHERE clause, or an aggregation.  Also, subqueries are undoubtedly one of the signs of the semantic complexity of queries, therefore we count the number of subqueries in each query.  We further consider the presence of JOINs since a JOIN means the need for additional source tables and that contributes to query complexity.  By counting the number of JOINS, we get a measure of how many data sources have to be combined to generate the final result.

The histograms for these three metrics applied to the three benchmarks are shown, respectively, in Figure\ref{fig:func-call}, \ref{fig:subquery}, \ref{fig:join}.  The distributions of function calls, subqueries and joins are all quite similar.  The gap between TPC-DS workload and the workloads of BIRD and SPIDER is even greater than that in the CTE metric above.  Most of TPC-DS queries have more than 5 function calls, some of them have more than 10.  In contrast, most of BIRD and SPIDER have 3 or fewer function calls, and never more than 5.  TPC-DS stands out as the only one among the three that makes regular use of subqueries.  The other two workloads appear to make minimal use of subqueries.  Just like subqueries and function calls, TPC-DS clearly outstrips BIRD and SPIDER in the number of JOINs.  This means that queries from TPC-DS almost always require data from a much larger number of tables in order to obtain the result.

Overall, TPC-DS queries are significantly more complex than BIRD and SPIDER queries in every metric we have considered as shown in the radar plot in Fig.~\ref{fig:radar}.


The comparison results are aggregated in Fig.~\ref{fig:radar} that shows the mean count of each metric of query complexity for the three benchmarks, normalized by the mean counts for TPC-DS.  TPC-DS clearly exhibits much greater query complexity in every metric.  Not only is its mean metrics higher than BIRD and SPIDER, they are multiple times higher in every metric.  This indicates that TPC-DS is a much more complex benchmark than BIRD and SPIDER, and that it is likely to be more challenging for AI models to generate queries based on TPC-DS workload.

\section{Evaluating LLMs Using TPC-DS Workload}
\label{sec:llm}

We have used 11 different LLMs to generate SQL queries based on the TPC-DS workload.  The LLMs are: gemini-1.5, gpt-4, codestral, mixtral-8x22b, mistral-large, llama3-70b, codellama-70b, llama3-8b, codellama-13b, codellama-35b, and codellama-7b.  We have used the following prompt to generate the queries.

\vspace{5mm}

\noindent {\bf System Prompt:}
\begin{tcolorbox}
\begin{verbatim}
You are a skilled PostgreSQL 
relational database developer.
Your task is to generate a PostgreSQL
query to answer user's question based
on given database schema without 
further explanations.

The database schema is:
{{ DDL_statements }}    
\end{verbatim}
\end{tcolorbox}

\vspace{5mm}

\noindent {\bf User Prompt:}
\begin{tcolorbox}
\begin{verbatim}
User question:
{{ question }}
\end{verbatim}
\end{tcolorbox}

For all the LLMs, we have encountered invalid queries generated by LLM, either with syntax error or containing incorrect schema information due to hallucination.  Our implementation utilizes the PostgreSQL database system to validate the generated queries. In case of invalid queries, we retry the generation process with a new user message containing the error message reported by PostgreSQL.

\vspace{5mm}

\noindent{\bf Additional User Prompt:}
\begin{tcolorbox}
\begin{verbatim}
Correct the query based on the error
message.

Query:
{{ sql }}

Error message:
{{ error_message }}
\end{verbatim}    
\end{tcolorbox}

\vspace{5mm}

The generation process is limited to maximum of three retries.  LLMs have different success rate after three retries.  The number of successful queries generated by each LLM is shown in Table~\ref{table:success}.  The LLMs gpt-4, gemini-1.5, and mistral-large have the highest success rate in generating queries based on the TPC-DS workload.  We can observe that the smaller models such as llama3-8b and codellama-34b or smaller have a very low success rate in generating queries.

\begin{table}[t]
\centering
\begin{tabular}{|l|c|}\hline
    LLM & Successful queries generated (out of 99)\\ \hline
    gpt-4 & 94 \\
    gemini-1.5 & 86 \\
    mistral-large & 75 \\
    codestral & 77 \\
    mistral-8x22b & 67 \\
    llama3-70b & 74 \\
    llama4-8b & 8 \\
    codellama-70b & 56 \\
    codellama-34b & 29 \\
    codellama-13b & 18 \\
    codellama-7b & 17 \\ \hline
\end{tabular}
\caption{Number of successfully generated queries based on TPC-DS query description.}
\label{table:success}
\end{table}

\subsection{Structural Complexity of Generated Queries}

In order to compare the structural complexity of the queries generated by the different LLMs, we again look at several count measures, such as number of columns and joins.  For each type of measure and each LLM, the mean of the counts in the queries is calculated, and then normalized over the mean count for TPC-DS queries.  These normalized mean counts are given in Table.~\ref{table:gen-table}.  We can see that gemini 1.5, gpt-4, mixtral and llama3-70b match the closest the structural complexity of TPC-DS.  The other LLMs generate markedly less complex queries.

It is encouraging that large LLMs are capable of generating queries that can match the complexity of TPC-DS workload.  But more importantly we want to evaluate the quality of these queries with respect to the gold queries given by TPC-DS.  In the next section, we will evaluate each of the generated queries using the bag-valued features as defined in Section~\ref{sec:features}.

\begin{figure}
    \includegraphics[width=0.8\textwidth]{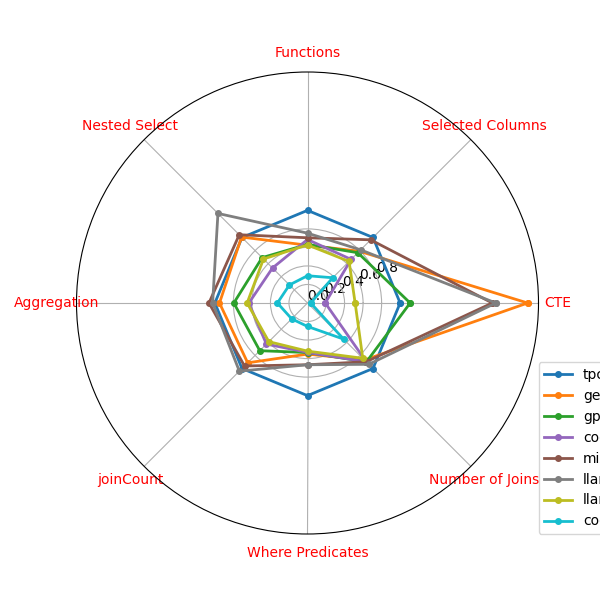}
    \caption{Comparison of generated queries}
    \label{fig:gen-radar-2}
\end{figure}

\subsection{Accuracy of Generated Queries}

We have defined 7 bag-valued features defined in Section~\ref{sec:features}.  Each feature is a function mapping SQL to a bag of discrete values.  To compare two queries: $Q$ the generated query by a LLM, and $Q^*$ the corresponding gold query given by TPC-DS benchmark, we use the Jaccard similarity coefficient between the two bags of values for each feature given by:
$$\mathrm{sim}_F(Q, Q^*) = \frac{|F(Q)\cap F(Q^*)|}{|F(Q)\cup F(Q^*)|}$$

Table~\ref{table:similarity} shows the mean similarity between the queries generated by 11 LLMs and the TPC-DS queries.  All LLMs are instructed with the instructions given in Section~\ref{sec:llm}.  Five of the top performing LLMs are visualized in a radar plot in Figure~\ref{fig:gen-radar} and Figure~\ref{fig:gen-radar-2}.  One can see that the accuracy of the generated queries is insufficient for practical real-world application.  In particular, none of the LLMs are able to generate queries that match the {\bf WHERE} predicates and {\bf JOIN} pairs used by TPC-DS.  This highlights the unique challenges posed by the TPC-DS workload in comparison with the BIRD and SPIDER benchmarks.

\begin{figure}
    \includegraphics[width=0.8\textwidth]{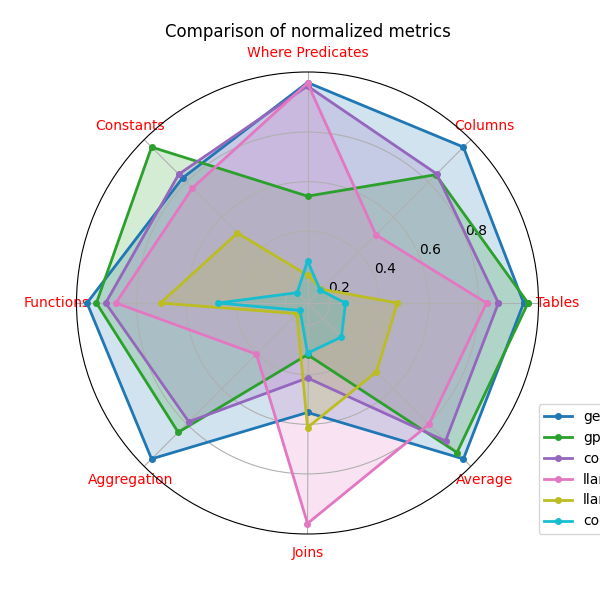}
    \caption{Comparison using normalized metrics}
\end{figure}

\section{Conclusion and Future Work}

We have compared TPC-DS with existing text-to-SQL benchmarks, BIRD and SPIDER, and found that TPC-DS queries exhibit a significantly higher level of structural complexity compared to the other two benchmarks.  We have also evaluated the performance of 11 LLMs in generating SQL queries based on the TPC-DS workload.  Our findings indicate that the current state-of-the-art generative AI models fall short in generating accurate decision-making queries.  The accuracy of the generated queries is insufficient for practical real-world application.  

Towards a satisfactory solution to complex SQL generation, we identify the following areas of future work.

\vspace{2mm}

\noindent {\bf Evaluating incremental SQL generation:} our experiments show the need of incremental SQL generation to improve the accuracy of generated queries.  This motivates us to investigate novel prompt strategies.  Based on the observation in Figure~\ref{fig:gen-radar}, we propose to generate WHERE clause and JOIN pairs as a separate LLM prompt, and use the results to prompt LLMs for the rest of the query.

\vspace{2mm}

\noindent {\bf Fine-tuning smaller models:} our experiments show that smaller models are not able to match the performance of larger models in the context of complex SQL generation.  In situations where cloud based LLMs are not feasible (due to cost or privacy concerns), we propose to investigate fine-tuning of smaller models to improve their performance.  In particular, smaller models can certainly be improved in terms of syntax and schema accuracy during generation.  Also, ensemble methods involving multiple fine-tuned LLMs can be used to combine the outputs of multiple smaller models to improve the overall accuracy.

\vspace{2mm}

\noindent {\bf Human-in-the-loop:} Complex SQL generation is a challenging task, and the burden of SQL generation should not be entirely on the AI model.  We propose to develop a novel human-in-the-loop workflow in which the AI model can identify the parts of the query that are difficult to generate, and prompt the user to provide additional information.

\bibliographystyle{unsrtnat}
\bibliography{references}

\end{document}